\documentclass[aps,pre,amsmath,amssymb,preprint,superscriptaddress]{revtex4}
\usepackage{graphicx}
\usepackage{color}
\begin{document}

\title{Dynamical Behavior Near a Liquid-Liquid Phase Transition in Simulations of Supercooled Water}

\author{Peter~H.~Poole}
\affiliation{Department of Physics, St. Francis Xavier University, Antigonish, Nova Scotia B2G 2W5, Canada}

\author{Stephen~R.~Becker}
\affiliation{Department of Physics, Wesleyan University, Middletown, CT 06459}
\altaffiliation[Present address: ]{Department of Applied \& Computational Math, California Institute of Technology, Pasadena, CA 91125}

\author{Francesco~Sciortino}
\affiliation{Dipartimento di Fisica and CNR-ISC, Universit\`a di Roma La Sapienza, Piazzale Aldo Moro 2, I-00185 Rome, Italy}

\author{Francis~W.~Starr}
\affiliation{Department of Physics, Wesleyan University, Middletown, CT 06459}

\begin{abstract} 
  We examine the behavior of the diffusion coefficient of the ST2 model
  of water over a broad region of the phase diagram via molecular
  dynamics simulations.  The ST2 model has an accessible liquid-liquid
  transition between low-density and high-density phases, making the
  model an ideal candidate to explore the impacts of the liquid-liquid
  transition on dynamics.  We locate characteristic dynamical loci in
  the phase diagram and compare them with the previously investigated
  thermodynamic loci.  The low-density liquid phase shows a crossover
  from non-Arrhenius to Arrhenius behavior, signaling the onset of a
  crossover from fragile-to-strong behavior.  We explain this crossover
  in terms of the asymptotic approach of the low-density liquid to a
  random tetrahedral network, and show that the temperature dependence
  of the diffusion coefficient over a wide temperature range can be
  simply related to the concentration of defects in the network.  Our
  findings thus confirm that the low-density phase of ST2 water is a
  well-defined metastable liquid.
\end{abstract}

\date{\today}
\maketitle

\section{Introduction}
\label{sec:intro}

It has long been appreciated that water, the most important of all
liquids, defies description in terms of simple liquid behavior in most
respects~\cite{eis-kauz,franks72}.  Many of the anomalies of the
thermodynamic and transport properties can be attributed to the hydrogen
bonds that dominate the intermolecular interactions~\cite{hes0, hes2,
  hes4, hes5, hes18, hes59}.  Of the many important studies of water
conducted over the last several decades, the 1992 proposal that a
liquid-liquid phase transition (LLPT) occurs in supercooled water has
had a particularly significant impact on water research~\cite{pses92}.
In this proposal, two phases of liquid water, a low density liquid (LDL)
and a high density liquid (HDL), become distinct below a critical point
located in the supercooled regime of the phase diagram.  H.~E.~Stanley
and co-workers have pursued the implications of this proposal over the
last 20 years~\cite{hes8, hes9, pess93, hes11, hes12, hes13, hes14,
  hes15, ms98, hes19, hes20, hes24, hes26, hes28, hes29, sas, hes33,
  hes34, hes38, hes40, hes43, hes45, hes46, hes48, hes51, hes55, hes58,
  hes60, hes62}, and much has been learned about the impact of a
liquid-liquid transition on the properties of water and related
systems~\cite{ms98review,hes32}.

The LLPT proposal remains controversial because its confirmation via
experiments on the bulk liquid has been thwarted by rapid ice nucleation
at the conditions at which the critical point is predicted to occur;
bulk studies of the ice melting lines provide indirect evidence of a
transition~\cite{ms98}.  The central strength of the LLPT proposal is
that it rationalizes the thermodynamical and dynamical anomalies of the
supercooled liquid while, at the same time, it accounts for the
occurrence in experiments of two widely different forms of amorphous
solid water (low density and high density amorphous ice) as the
sub-glass-transition manifestations of the LDL and HDL
phases~\cite{mishima85,pess93,mishima94,Angell:2004p1313,hes34,hes40,Loerting:2006p4014}.
Indeed, the possibility of a LLPT has now been investigated across the
entire class of liquids in which tetrahedral bonding dominates the local
structure.  This class of systems includes water,
Si~\cite{Sastry:2003p3801}, SiO$_2$~\cite{wac97}, and
BeF$_2$~\cite{Hemmati:2001p5433}, as well as nanoparticle liquids
tailored to exhibit tetrahedral
interactions~\cite{Hsu:2008p5245,Dai:2010p7264}.  Additionally, it has
also been shown that liquids with symmetric interactions, but with a
competition between low density and high density packings, may also
exhibit LLPT behavior~\cite{hes26, hes33, hes38, Kumar:2005p7317}.

Due to the challenges imposed by crystallization on experiments of
supercooled water, computer simulations have played a central role in
the development of the LLPT proposal for water, and other tetrahedral
liquids.  While a LLPT occurs in a variety water models, one of the most
accessible and clearest examples is in the venerable ST2
model~\cite{ST2}, one of the earliest point-charge models for water.
``ST2 water'' has been extensively studied to clarify the nature of such
a LLPT, in particular with regard to thermodynamic and structural
properties~\cite{pses92,pses93,spes97, hes13,pg99, Poole:2005p2770,
  lpd09, Cuthbertson:2011p7361}.  While much is known about the dynamics
of ST2 water~\cite{sf89, sgs92, hes5, hes6, hes7,pg99, Becker:2006p15,
  hes51} a comprehensive study of the dynamical properties comparable in
scope to the thermodynamic studies is lacking.  Therefore, in the
present work, we focus on the dynamical properties of the ST2 model over
a wide range of states that encompasses the vicinity of the LLPT.  We
show that there are striking differences in the nature of the dynamics
of the HDL and LDL phases.  The LDL phase presents a particular
challenge, as the relaxation time of the liquid increases extremely
rapidly with decreasing temperature $T$.  However, our results
demonstrate that the equilibrium dynamical properties of the LDL phase
can be understood from the behavior observed in the region accessible to
our simulations.  These results emphasize the central role of the
developing network of hydrogen bonds for understanding the behavior of
the liquid, especially of the LDL phase.

\section{Simulations}
\label{sec:sims}

Our data are generated via molecular dynamics simulations of the ST2
potential~\cite{ST2}.  The simulations follow the same protocol as those
in Ref.~\cite{Poole:2005p2770}.  Specifically, we attempt simulations at
thermodynamic state points in the density range $\rho = 0.80$~g/cm$^3$
to 1.20~g/cm$^3$ at intervals $\Delta \rho = 0.01$~g/cm$^3$, and the
temperature range $T=200$~K to 400~K at intervals of $\Delta T = 5$~K.
At each state point considered, we simulate $N=1728$ molecules, unless
noted otherwise.  We are able to equilibrate the liquid at all chosen
densities for $T\ge255$~K; for $T<255$~K, the lowest density studied is
limited by the extreme length of simulation required. In total, we
examine more than 1500 different state points.  The configurations used
to start simulations are the final configurations from
Ref.~\cite{Poole:2005p2770}, where a detailed equation of state (EOS) of
ST2 water was evaluated.  Each simulation is run until the mean-squared
displacement (MSD) of oxygen atoms reaches 1~nm$^2$ (roughly three
interparticle spacings), much longer than needed to observe diffusive
behavior.  We use an integration time step of 1~fs and employ the
Berendsen heat bath with a time constant of 2~ps to control $T$ during
the production run, in order to compensate for a possible minor drift of
the energy in the very long simulation runs.  The electrostatic
potential is truncated at 7.8~\AA~and the energy and pressure are
corrected using the reaction field method~\cite{allen-tildesley};
periodic boundaries are used to minimize size effects.  Trajectory
information was written to disk every 100~fs for $T\ge250$~K, and every
10~ps for $T<250$~K.

\begin{figure}[t]
\includegraphics[width=13cm]{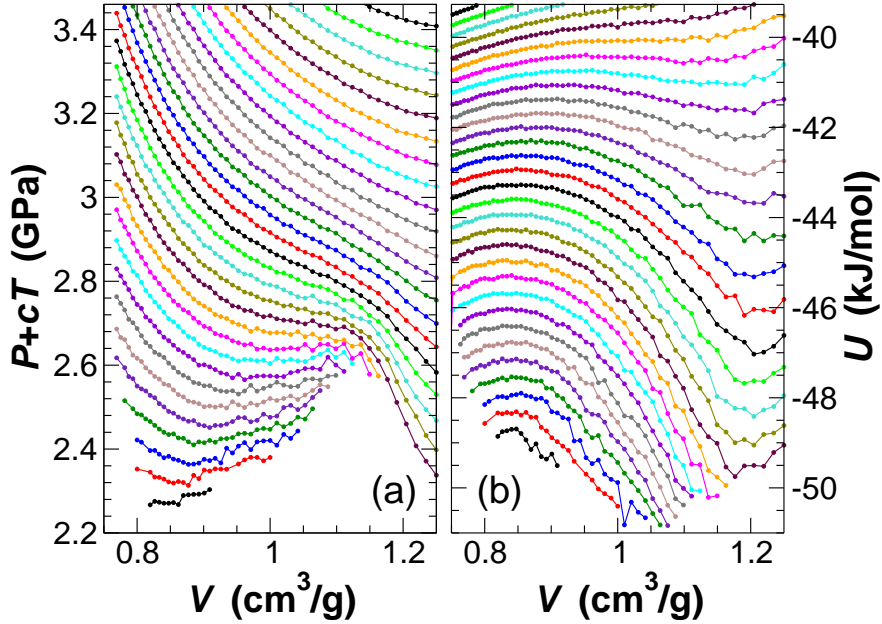}
\caption{(a) Pressure $P$ and (b) potential energy $U$ as a function of
  volume $V$ along isotherms for ST2 water, from our $N=1728$
  simulations.  In both panels, we show isotherms from $T=200$ to
  $350$~K, in 5~K steps, from bottom to top.  In (a), each isotherm is
  shifted by $cT$, with $c=10$~MPa/K, in order to facilitate comparison
  of the curves.  In (b), the minima of $U$ occurring at low $T$ and in
  the vicinity of $V=1.2$~cm$^3$/g ($\rho=0.83$~g/cm$^3$) identify
  conditions at which a particularly well developed random tetrahedral
  network occurs in the liquid.}
\label{fig:thermo}
\end{figure}

Figure~\ref{fig:thermo} summarizes the thermodynamic properties of the ST2
liquid in the region examined here.  Fig.~\ref{fig:thermo}(a) shows
the pressure $P$ as a function of volume $V$ along isotherms.  As $T$
decreases, these isotherms first inflect, and then become progressively
flatter, leading to the realization of a liquid-liquid critical point at
which $(\partial P/\partial V)_T=(\partial^2 P/\partial V^2)_T=0$.  The
critical point conditions in ST2 water have been estimated to occur at
$T_c=247\pm 3$~K, $P_c=185\pm 15$~MPa, and $\rho_c=0.955\pm
0.01$~g/cm$^3$~\cite{lpd09,Cuthbertson:2011p7361}.

Figure~\ref{fig:thermo}(b) presents the variation of the potential energy $U$
as a function of $V$ along isotherms of the liquid.  As noted in
previous work~\cite{spes97,pwmnoi}, the emergence with decreasing $T$ of
significant negative curvature in these isotherms is a thermodynamic
signature of the approaching liquid-liquid instability: To be stable
with respect to phase separation, the Helmholtz free energy $A$ must
have positive curvature $(\partial^2 A/\partial V^2)_T>0$.  Since
$A=U-TS$ (where $S$ is entropy), the negative curvature shown in
Fig.~\ref{fig:thermo}(b) indicates an energetic driving force for phase
separation that becomes dominant for $T<T_c$~\cite{spes97}. It is also
interesting to observe the minima of $U$ occurring at low $T$ and in the
vicinity of $V=1.2$~cm$^3$/g ($\rho=0.83$~g/cm$^3$). The presence of
this minimum has also been observed in several other models of
tetrahedral liquids\cite{spes97, pwmnoi, silicaCristiano, floatingbond}
and signals the optimal network volume (or density), {\it i.e.}\ the
conditions at which a particularly well developed random tetrahedral
network occurs in the liquid.  Indeed, hydrogen bonding requires a
well-defined distance and orientation to be effective, and at the
optimal density (which is usually close to the fully bonded open crystal
density) geometric constraints allow the establishment of a fully bonded
disordered network.

\begin{figure}
\includegraphics[width=8cm]{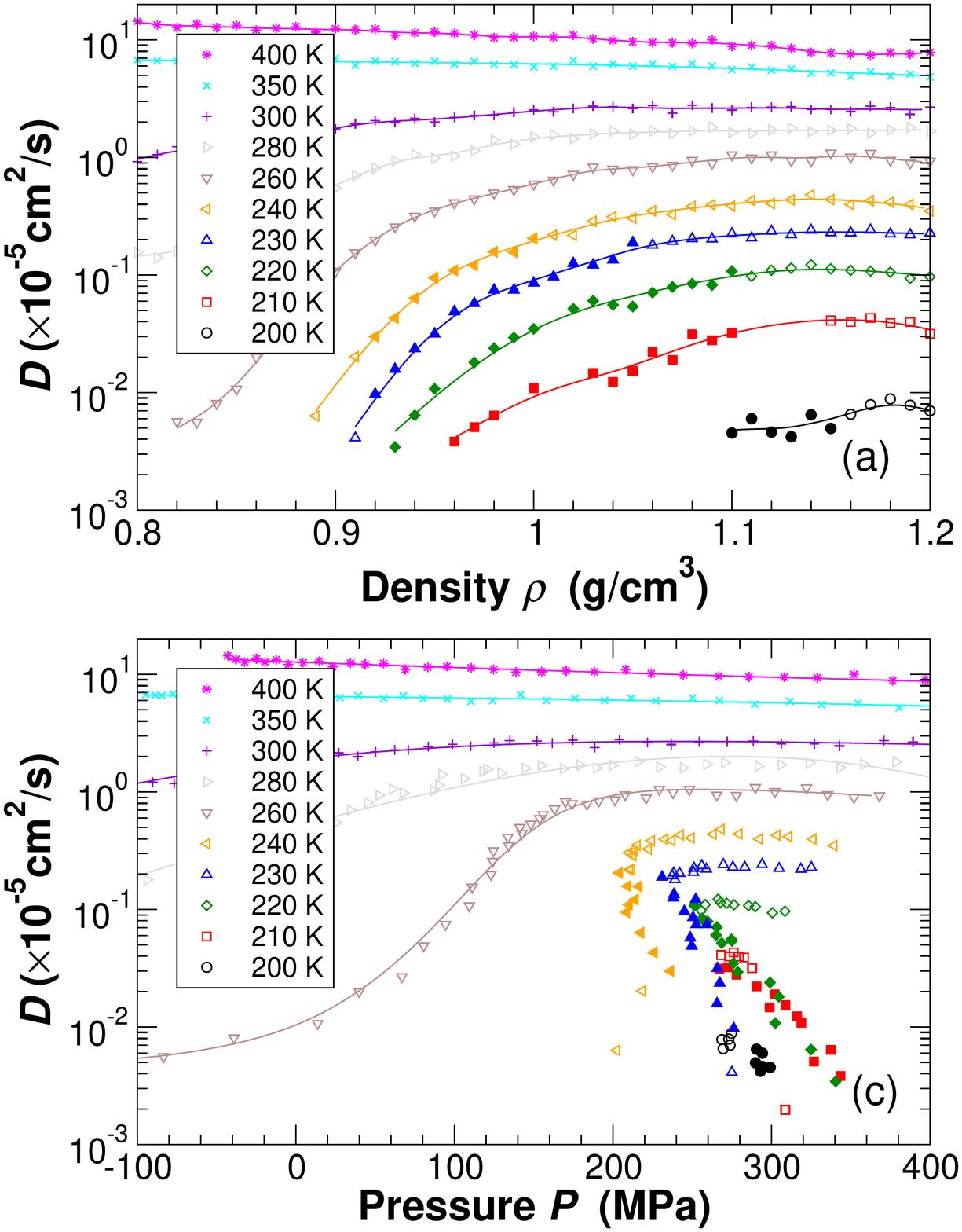}
\includegraphics[width=8cm]{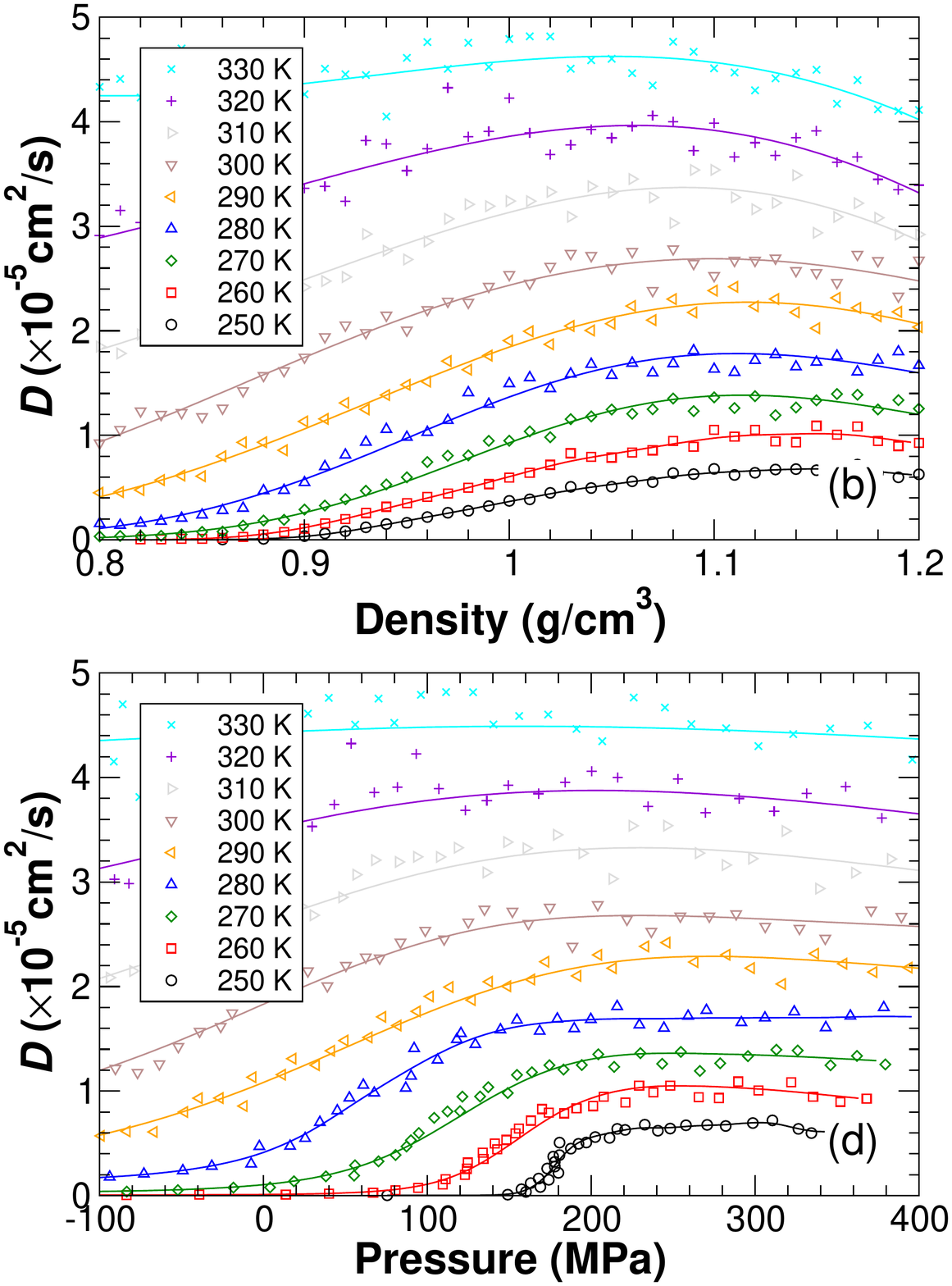}
\caption{ The diffusion coefficient $D$ along isotherms as a function of
  density $\rho$ ((a) and (b)) or pressure $P$ ((c) and (d)).  The lines
  are intended only as a guide for the eye.  We plot only a selection of
  our available isotherms to avoid overcrowding in the figure. We show
  both log ((a) and (c)) and linear ((b) and (d)) scales to cover a
  broad range of $D$ (log scale) and to show the form of the maxima of
  $D$ (linear scale).  The solid symbols indicate state points in the
  unstable zone of the liquid-liquid coexistence region.  Note that $D$
  is a multi-valued function of $P$ along isotherms for $T<250$~K
  because $P$ is a non-monotonic function of $\rho$ in the co-existence
  region along these isotherms.}
\label{fig:D-isotherms}
\end{figure}
  
\begin{figure}[!t]
\includegraphics[width=10cm]{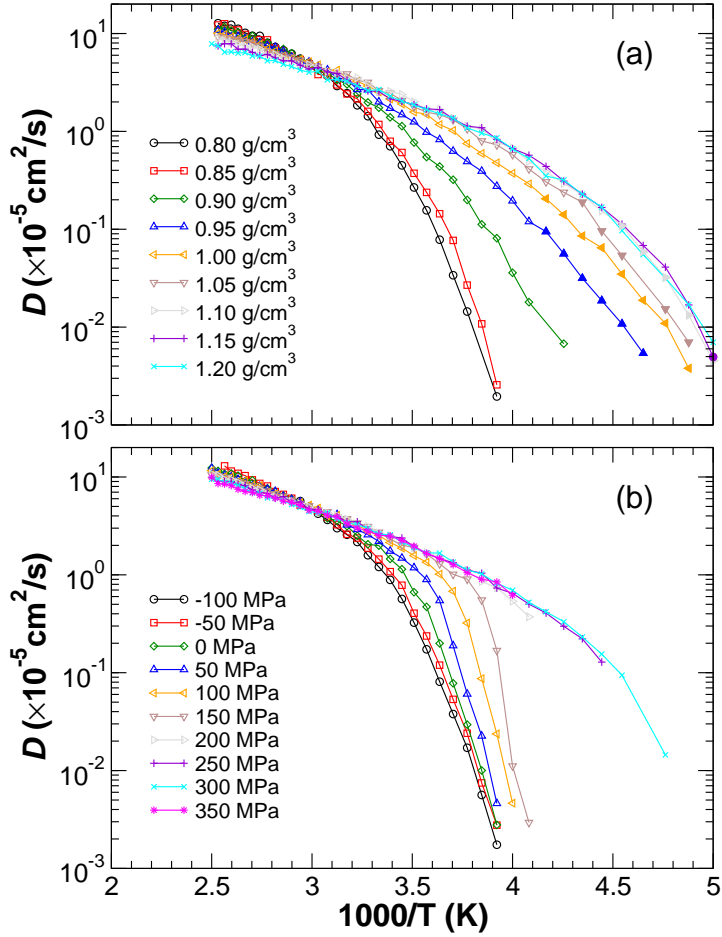}
\caption{Diffusion coefficient $D$ as a function of $T$ along (a)
  isochores and (b) isobars.  Along isochores the solid symbols indicate
  systems from the unstable zone of the coexistence region.  No data for
  unstable states are shown along the isobars in (b), since $D$ is a
  multi-valued function of pressure in this region
  [see~Fig.~\ref{fig:D-isotherms}(c,d)].  We show only a selection of
  densities and pressures to avoid overcrowding the figure.}
\label{fig:D-temperature}
\end{figure}

\begin{figure}[t]
\includegraphics[width=8cm]{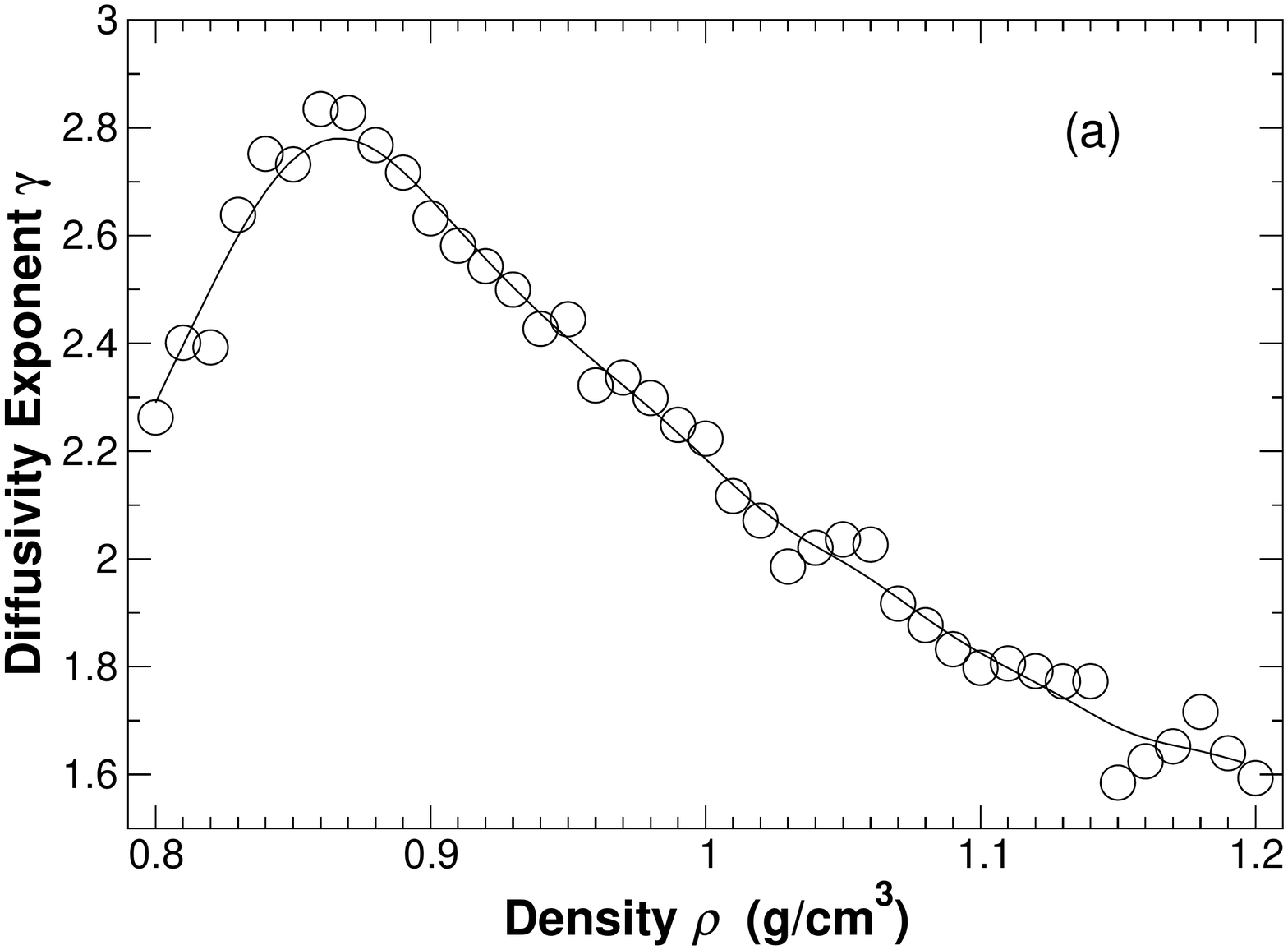}
\includegraphics[width=8cm]{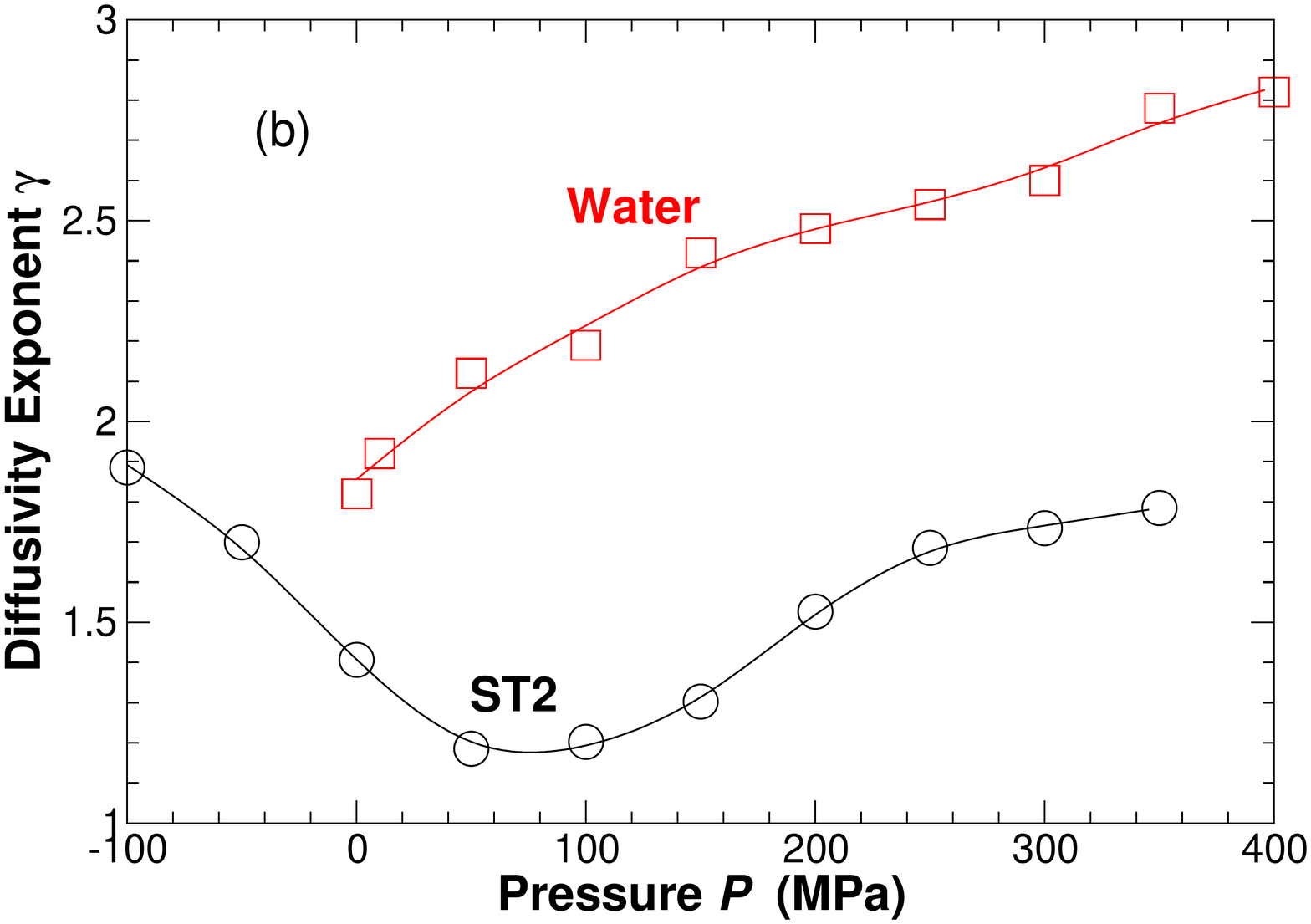}
\caption{The diffusivity exponent $\gamma$ as a function of (a) density
  and (b) pressure.  The pressure dependence of $\gamma$ parallels that
  of water in the region where experimental data is available.
  Experimental data are from Ref.~\cite{prielmeier,prielmeier2}.  The
  lines are only a guide to the eye.}
\label{fig:gamma}
\end{figure}
 
\begin{figure}
\includegraphics[width=8cm]{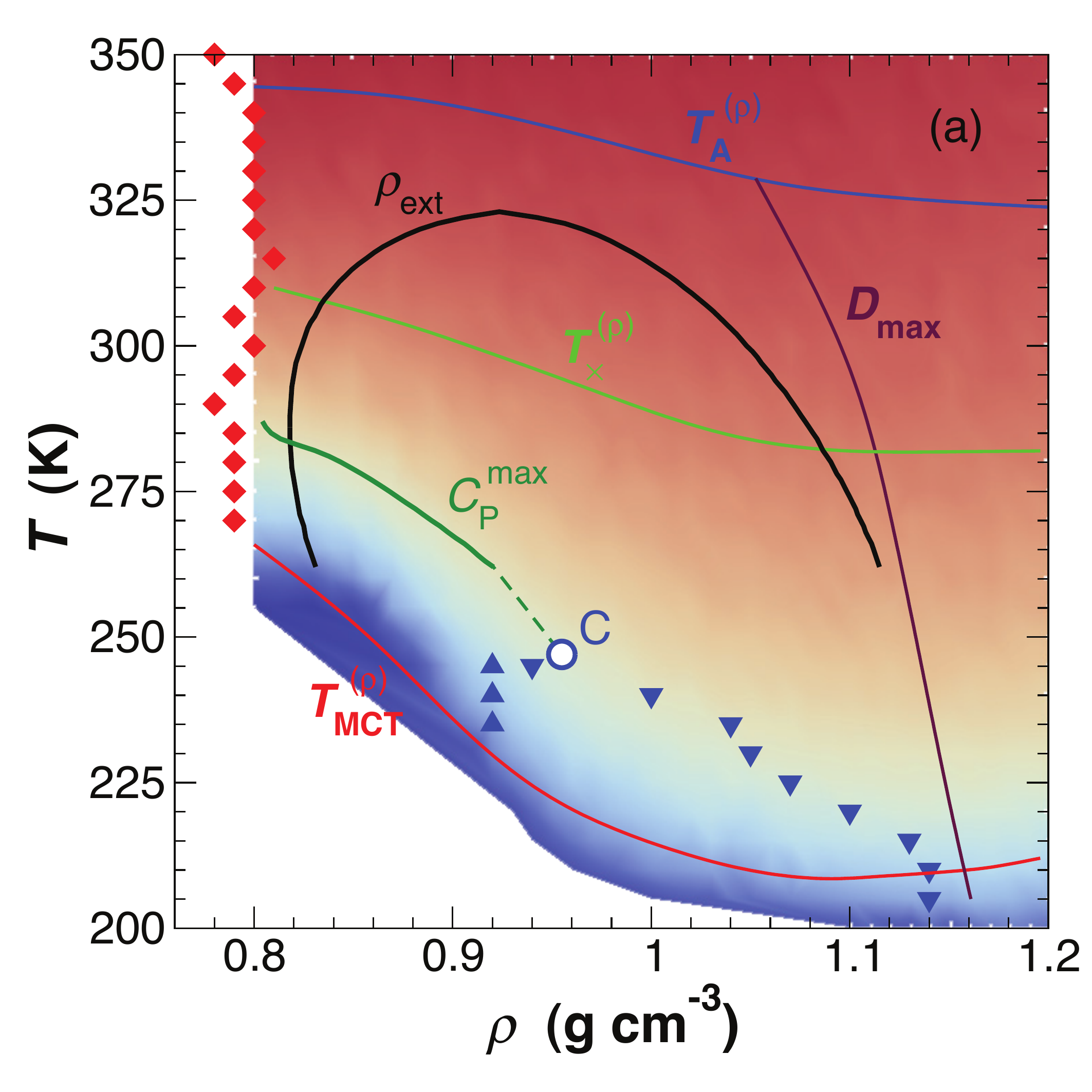}
\includegraphics[width=8cm]{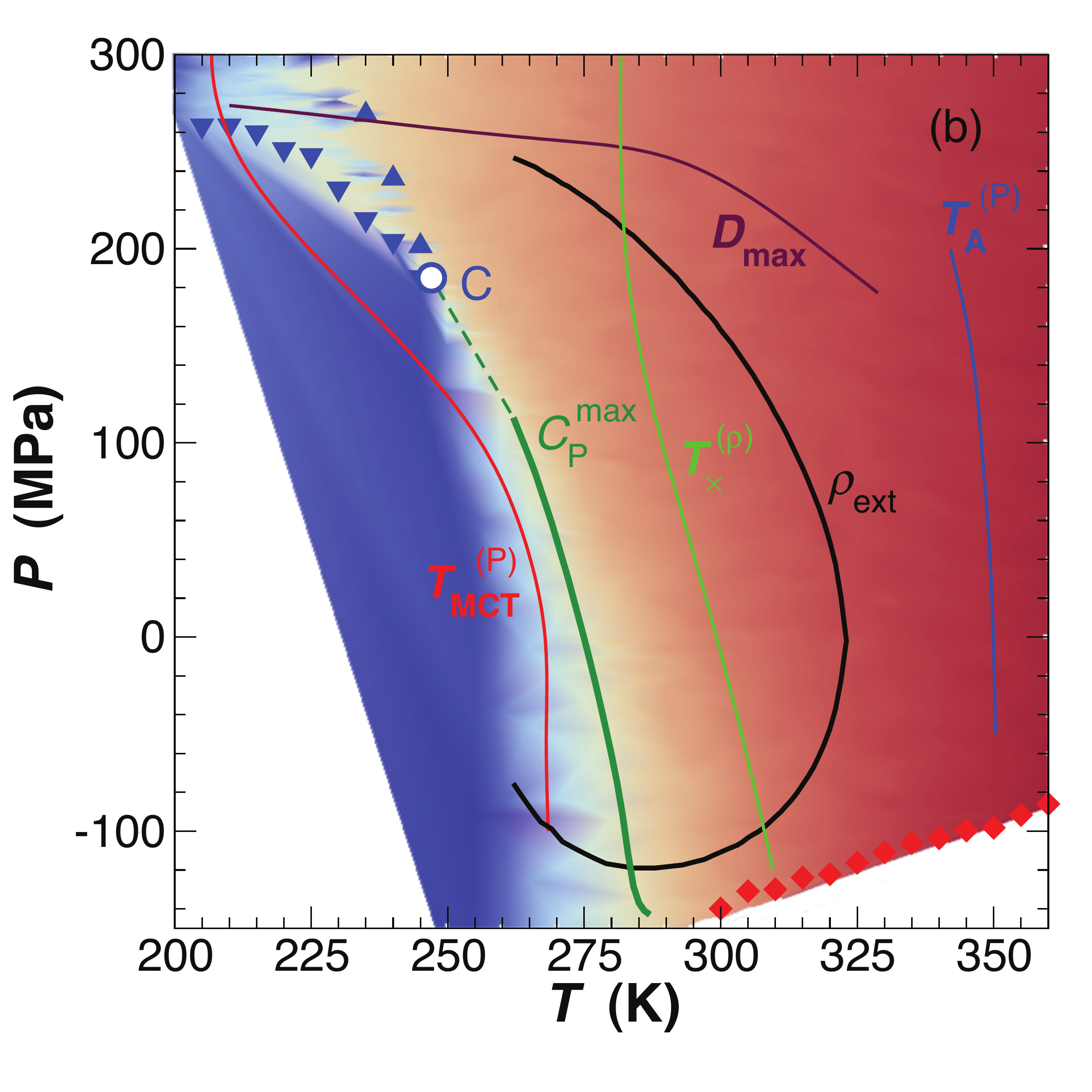}
\caption{The (a) $T-\rho$ and (b) $P-T$ phase diagrams, combined with
  the dynamical features determined here: $D_{\rm max}$, the locus of
  extrema of $D$ along isotherms; $T_A^{(\rm X)}$, the onset of
  non-Arrhenius behavior; $T_\times^{\rm X} $, the breakdown of
  Stokes-Einstein behavior (from \cite{Becker:2006p15,Kumar:2007p162});
  $T_{\rm MCT}^{\rm X}$, the extrapolated mode-coupling divergence
  temperature.  The superscript $X$ for the last three loci indicates
  whether the locus was determined along isochores ($X=\rho$) or isobars
  ($X=P$).  The color map indicates the value of $\log D$; red (blue)
  represents the highest (lowest) values of $D$.  The liquid-gas
  spinodal is indicated by red diamonds; the liquid-liquid spinodals are
  indicated by blue triangles.  The liquid-liquid critical point
  location is indicated by the open circle.  The locus of specific heat
  maxima $C_P^{\rm max}$ tracks the maxima of $C_P$ along isobars; the
  dotted portion of the line is an extrapolation to the critical point,
  where $C_P$ diverges.  $\rho_{\rm ext}$ labels the locus of density
  extrema; inside the $\rho_{\rm ext}$ locus, the isobaric expansivity
  is negative, while outside it is positive.  All thermodynamic data are
  from Ref.~\cite{Poole:2005p2770}. }
\label{fig:Trho-diagram}
\end{figure}

 \section{Diffusivity of the ST2 Model}
\label{sec:diffusion}

For each state point simulated, we calculate the mean-squared
displacement $\langle r^2(t) \rangle$ and evaluate the diffusion
coefficient from the Einstein relation
\begin{equation} 
D =\lim_{t\rightarrow \infty} \frac{\langle r^2(t) \rangle}{6t}.
\end{equation}
Here $\langle ... \rangle$ represents an average over possible time
origins.  Since we have more than 1500 state points to consider, we
automate the evaluation of $D$ by a linear fit of $\langle r^2(t)
\rangle$ for all data such that $\langle r^2(t)\rangle>0.5$~nm$^2$, a
restriction that ensures that all fitted data are well within the
diffusive regime.

We first consider $D$ along isotherms, as shown
in~Fig.~\ref{fig:D-isotherms}.  For $T<335$~K, $D$ exhibits a (weak) maximum
with increasing $\rho$ or $P$, as known experimentally~\cite{prielmeier,
  prielmeier2}. This feature is normally attributed to the breaking of
hydrogen bonds with increasing density, which allows for increased
diffusion, until packing considerations become dominant and $D$
decreases.  As compared with water, the ST2 model overestimates the
pressure of the maximum in $D$, which is not surprising given that ST2
overemphasizes the tetrahedral structure relative to
water~\cite{hpss97}.  Hence, there exists a locus of points (which we
denote $D_{\rm max}$) on the EOS surface at which $D$ is a maximum along
isotherms; the shape of this locus in the phase diagram is discussed
below.  Fig.~\ref{fig:D-isotherms} also includes data obtained for $T<T_c$.
The filled data points indicate simulations in the unstable regime where
our system (simulated at constant volume) is phase separated into
regions of LDL and HDL.  The values of $D$ in this region thus reflect a
weighted average over the LDL and HDL phases.  We note that $D$
decreases by nearly two orders of magnitude as the system progressively
transforms from pure HDL to pure LDL along the lowest $T$ isotherms.

We next present the $T$-dependence of $D$ along isochores and isobars
in Fig.~\ref{fig:D-temperature}.  At high $T$, $D$ is described by the
expected Arrhenius behavior
\begin{equation}
D = D_0 \exp [-E/k_BT],
\label{eq:arrhenius}
\end{equation}
where $E$ is the activation energy for diffusion and $D_0$ is the
limiting high-$T$ diffusion coefficient (both determined from fitting
eq.~\ref{eq:arrhenius} to the data).  On cooling below a temperature
$T_A$, $D$ exhibits so-called ``super-Arrhenius'' behavior, where $D$
decreases faster than expected relative to the high-$T$ behavior.  This
rapid decrease is typical of glass-forming liquids as they approach the
glass transition temperature $T_g$.  We estimate $T_A$ for $D$ along
either isochores or isobars by finding that $T$ at which $k_B T/E\; \ln
(D/D_0) > 1.02$; by construction, this quantity must equal unity for
high-$T$ Arrhenius behavior~\cite{sds98}.

The non-Arrhenius behavior of $D$ for many glass-forming liquids is
well-accounted for by the mode-coupling theory (MCT) for the glass
transition\cite{goetze}, which predicts that
\begin{equation}
D \sim (T-T_{\rm MCT})^{\gamma},
\label{eq:mct}
\end{equation}
where $T_{\rm MCT}$ is the temperature of an avoided vitrification of
the ideal MCT.  Although discussion continues concerning the region of
validity of the MCT approach, the phenomenological appearance of
power-law behavior immediately below $T_A$ is widely acknowledged.
Moreover, power-law behavior of $D$ in water at atmospheric pressure has
been appreciated since the 1970's~\cite{angell82}.  This power-law
behavior of $D$ in water was first interpreted as possible evidence for
an underlying spinodal singularity, but more recently it has been
connected with the MCT approach\cite{gstc96,sgtc96,sss,sss2}.
Accordingly, we fit separately the isobaric and isochoric data for $D$
to eq.~\ref{eq:mct} to evaluate the locus of $T_{\rm MCT}$ in the phase
diagram, which we discuss below.  Fits must be considered along both
isochoric and isobaric paths, since the diffusivity exponent $\gamma$
will depend on the path of the approach to $T_{\rm MCT}$.  Additionally,
care must be taken when making this fit, since it is apparent
from~Fig.~\ref{fig:D-temperature} that it appears $D$ may return to Arrhenius
behavior at lower $T$, a phenomenon which we will discuss in detail in
the following section.  Therefore, we have excluded data where the
behavior may revert to Arrhenius.

We find that the diffusivity exponent $\gamma$ is non-monotonic with
density, reaching a maximum of $\approx 2.8$ at $\rho \approx
0.88$~g/cm$^3$ (Fig.~\ref{fig:gamma}).  The exponent $\gamma$ is also
non-monotonic as a function of pressure.  A comparison with experimental
data for $\gamma$ shows that the behavior of $\gamma$ for ST2 is roughly
parallel in the region where both experimental and simulation data are
available.  This is in contrast with the SPC/E model of water, where the
opposite pressure dependence occurs~\cite{sss2}.

To summarize the properties of $D$, we collect the resulting
characteristic features from the isothermal, isochoric, and isobaric
plots (Fig.~\ref{fig:D-isotherms} and Fig.~\ref{fig:D-temperature}) and plot them
together with the known thermodynamic features. We show these features,
along with a colormap for $D$, in the $\rho-T$ plane and in the $P-T$
plane in Fig.~\ref{fig:Trho-diagram}.  As expected from
Ref.~\cite{Errington:2001p7324}, the locus $D_{\rm max}$ lies outside
the region of negative entropy-volume correlations, delineated by the
locus of extrema of density $\rho_{\rm ext}$.  The existence of a
maximum in $D$ also results in non-monotonic behavior of $T_A$ and
$T_{\rm MCT}$, since these temperatures represent a nearly constant
value of $D$.  The locus of $T_{\rm MCT}$ is nearly coincident with the
lower bound of our simulated data, since this also represents the time
scale where simulations become prohibitively lengthy.  We also show the
locus $T_\times$ of the breakdown of the Stokes-Einstein (SE)
relationship from Ref.\cite{Becker:2006p15,Kumar:2007p162}.  It is
interesting to note that $T_\times>T_{\rm MCT}$, since the breakdown of
the SE relation is often linked with $T_{\rm MCT}$.  For low pressure,
$T_\times$ is correlated with (but not coincident with) the extrema of
the specific heat~\cite{Kumar:2007p162}.  Finally, the shape of the loci
$T_A$, $T_{\rm MCT}$ and $T_X$ are all similar, with the non-monotonic
behavior becoming more pronounced with lower $T$.

\section{Structure and Dynamics of the Low Density Liquid Phase}
\label{sec:frag-strong}

One of the most challenging aspects of the liquid-liquid phase
transition in ST2 water is characterizing the properties of the LDL
phase.  As shown in the previous section, at low density ($\rho<\rho_c$
or $P<P_c$) the value of $D$ decreases far more rapidly as $T$ decreases
than for $\rho>\rho_c$.  Obtaining equilibrium values for both
structural and dynamical properties of the LDL phase is therefore
particularly demanding of computational resources.  In this section, we
examine the behavior of the liquid in this low density regime, and show
that many properties of the LDL phase can be determined from the
behavior observed in the region accessible to our simulations.

In particular, we focus on states along the optimal density
$\rho=0.83$~g/cm$^3$ isochore.  As shown in Fig.~\ref{fig:thermo}(b), this
density approximately corresponds to a minimum in isotherms of $U$
versus $V$, indicating that the structure of the random tetrahedral
network (RTN) is particularly well developed at this density.  We
therefore expect that the characteristic properties of the LDL phase
will be most prominent at this density.

To complement the $N=1728$ simulation results described in the previous
section, we examine an ensemble of 40 independently initialized and
equilibrated simulations along the $\rho=0.83$~g/cm$^3$ isochore for a
system of size $N=216$. The smaller system size allows us to probe
longer time scales than for $N=1728$.  In addition, by averaging our
results at each $T$ over the 40 independent runs, we can significantly
reduce the statistical error of our results.  We thus use these smaller
systems to carefully parse the behavior of the low-density liquid.

To evaluate the dynamical behavior of the $N=216$ system at
$\rho=0.83$~g/cm$^3$, in each of the 40 runs, the diffusion coefficient
is estimated from $D=\langle r^2 \rangle/6t$, where $\langle r^2
\rangle$ is the mean squared displacement at the end of the run, and $t$
is the time of the run.  All our production runs for $N=216$ are carried
out until $\langle r^2 \rangle=1.0$~nm$^2$, or $t=0.5$~ns, which ever
takes longer to achieve.  At the lowest $T$ (255~K), the longest runs
require up to 350~ns.  We then average the results over the 40 runs.

\begin{figure}[!t]
\includegraphics[width=12cm]{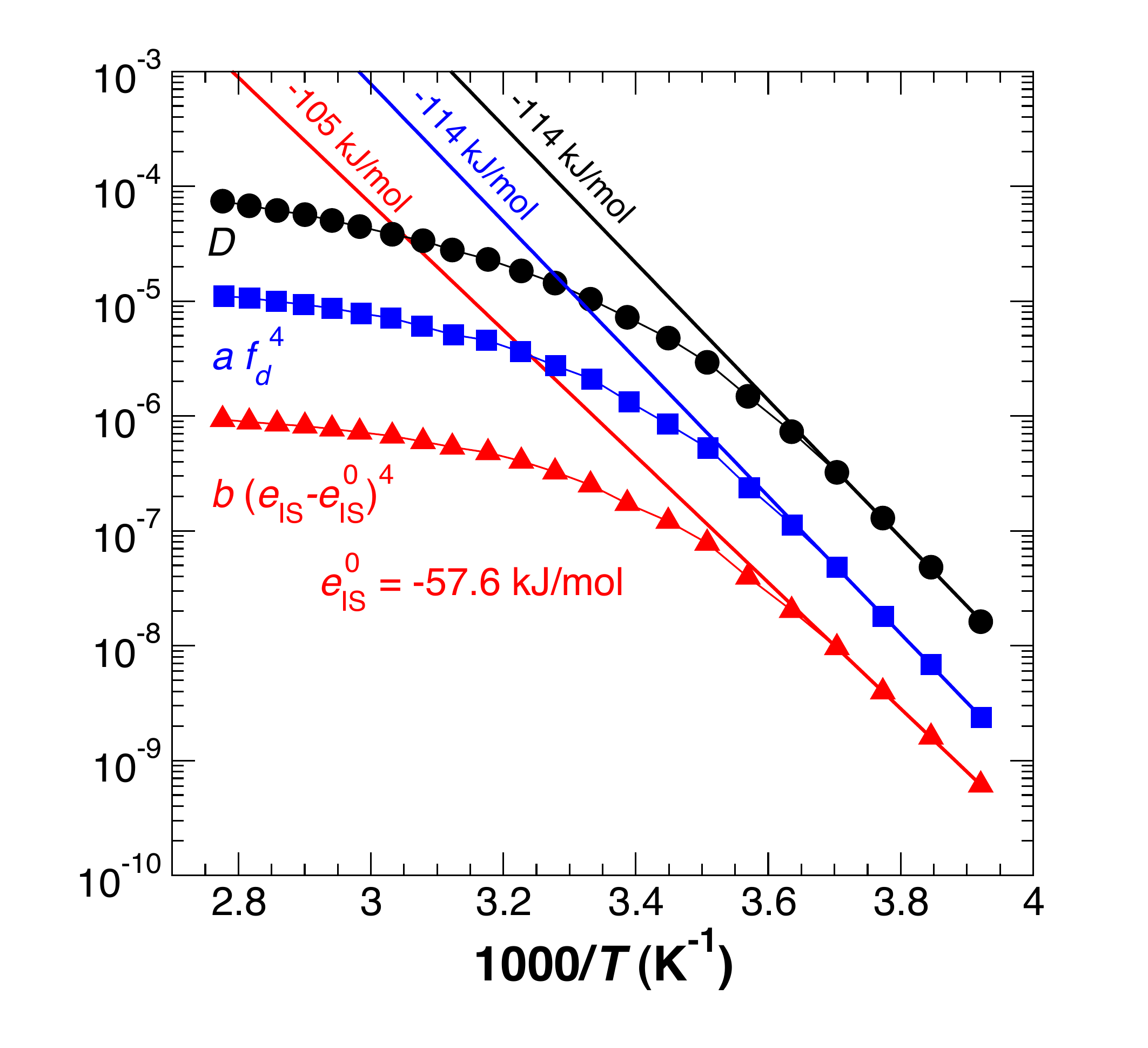}
\caption{Arrhenius plots of $D$, $a\cdot f_d^4$, and $b\cdot (e_{\rm
    IS}-e_{\rm IS}^0)^4$;  $f_d$ is the fraction of defects found in
  instantaneous liquid configurations.  All curves are for liquid ST2
  water along the $\rho=0.83$~g/cm$^3$ isochore, as determined from our
  $N=216$ simulations.  $D$ is plotted in cm$^2$/s, while $f_d^4$ and
  $(e_{\rm IS}-e_{\rm IS}^0)$ are multiplied by arbitrary constants $a$
  and $b$ to facilitate comparison within the same plot.  The straight
  lines are fits to an expression proportional to $\exp(-E_a/RT)$ for
  the four lowest $T$ points along each curve; each of these lines is
  labeled by the activation energy $E_a$ obtained from the fit.}
\label{fig:Dfe}
\end{figure}

\begin{figure}[!t]
\includegraphics[width=10cm]{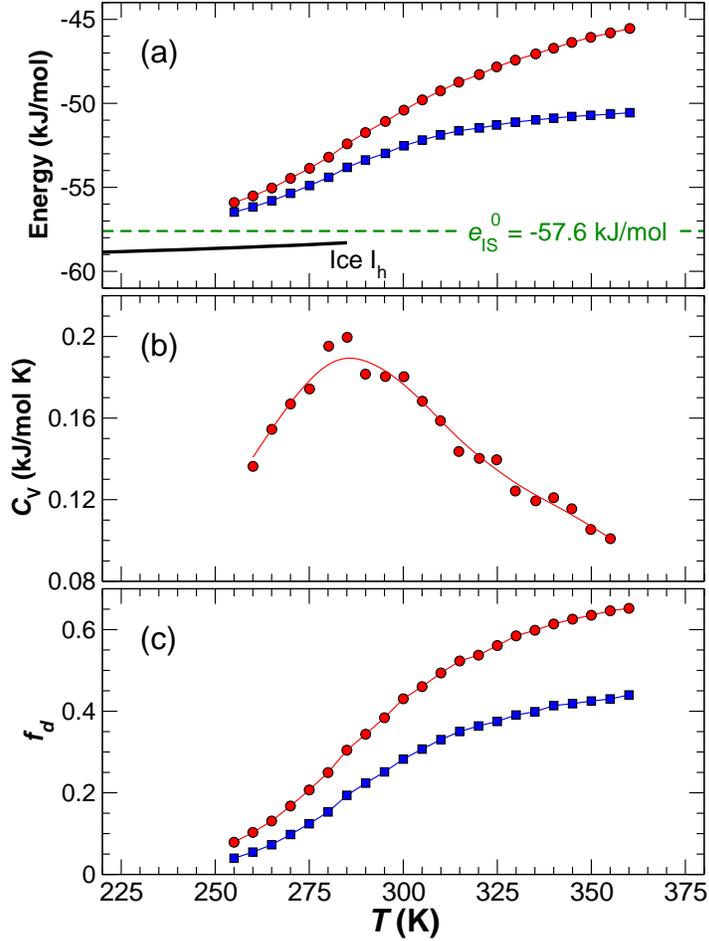}
\caption{Behavior of liquid ST2 water as a function of $T$ along the
  $\rho=0.83$~g/cm$^3$ isochore, determined from the $N=216$
  simulations.  (a) $U-3RT$ for the liquid (circles), and for
  crystalline ice I$_h$ (solid line), compared with the behavior of
  $e_{\rm IS}$ (squares) from the inherent structures.  The dotted green
  line shows the estimated limiting $e_{\rm IS}^0$ for amorphous states,
  larger than the ice energy. (b) The isochoric heat capacity
  $C_V$ of the liquid.  The solid line is only a guide for the eye. (c)
  The fraction of defects $f_d$ as found from instantaneous (circles)
  and inherent structure (squares) configurations of the liquid.}
\label{fig:huge}
\end{figure}

The results for $D$ as a function of $T$ are shown in an Arrhenius plot
in Fig.~\ref{fig:Dfe}.  Consistent with the results of
Fig.~\ref{fig:D-temperature}, we find that the $T$-dependence of $D$ crosses
over from non-Arrhenius (fragile) at higher $T$ to Arrhenius at the
lowest $T$ with an activation energy $E=114$~kJ/mol.  This appears to be
the beginning of a ``fragile-to-strong crossover'' (FSC); such
behavior has been observed and studied in a number of systems in which
RTN-like structure emerges at low $T$
\cite{sss,sas,orbach,SaikaVoivod:2001p5119,pwmnoi,silicaCristiano,Xu:2005p4066}.
Since we have not yet reached the low energy RTN, the activation energy
has not reached its low-$T$ asymptotic value.  The approach to the RTN
can exhibit intermediate Arrhenius behavior over several decades in $D$
before reaching the asymptotic limit~\cite{sas}.

To support our interpretation that we are approaching strong liquid
behavior as $T$ decreases, we test the expectation that the FSC is
associated with the approach to the lowest lying minima of the liquid's
potential energy landscape (PEL).  To confirm this behavior in ST2
water, we carry out conjugate gradient quenches of a large number of
configurations (at least 400 for each value of $T$ studied) in order to
evaluate the average energy of a liquid configuration when quenched to
its nearest local minimum of the PEL, referred to as the average
inherent structure energy, $e_{\rm IS}$.

In Fig.~\ref{fig:huge}(a) we show both $U$ and $e_{\rm IS}$ as a function of
$T$ for $\rho=0.83$~g/cm$^3$.  In both cases we find that an inflection
occurs in the vicinity of $T=285$~K, and that at lower $T$ the rate of
decrease of the energy slows with decreasing $T$.  Fig.~\ref{fig:huge}(a)
also shows $U$ for the ice I$_h$ crystal, the value of which sets a
lower bound on $e_{\rm IS}$ for the liquid.  The relative shape and
position of these curves suggests that the energy of the low density
liquid is approaching a low $T$ limit associated with the ``bottom'' of
the liquid PEL that lies above the crystal state.

\begin{figure}[!t]
\includegraphics[width=12cm]{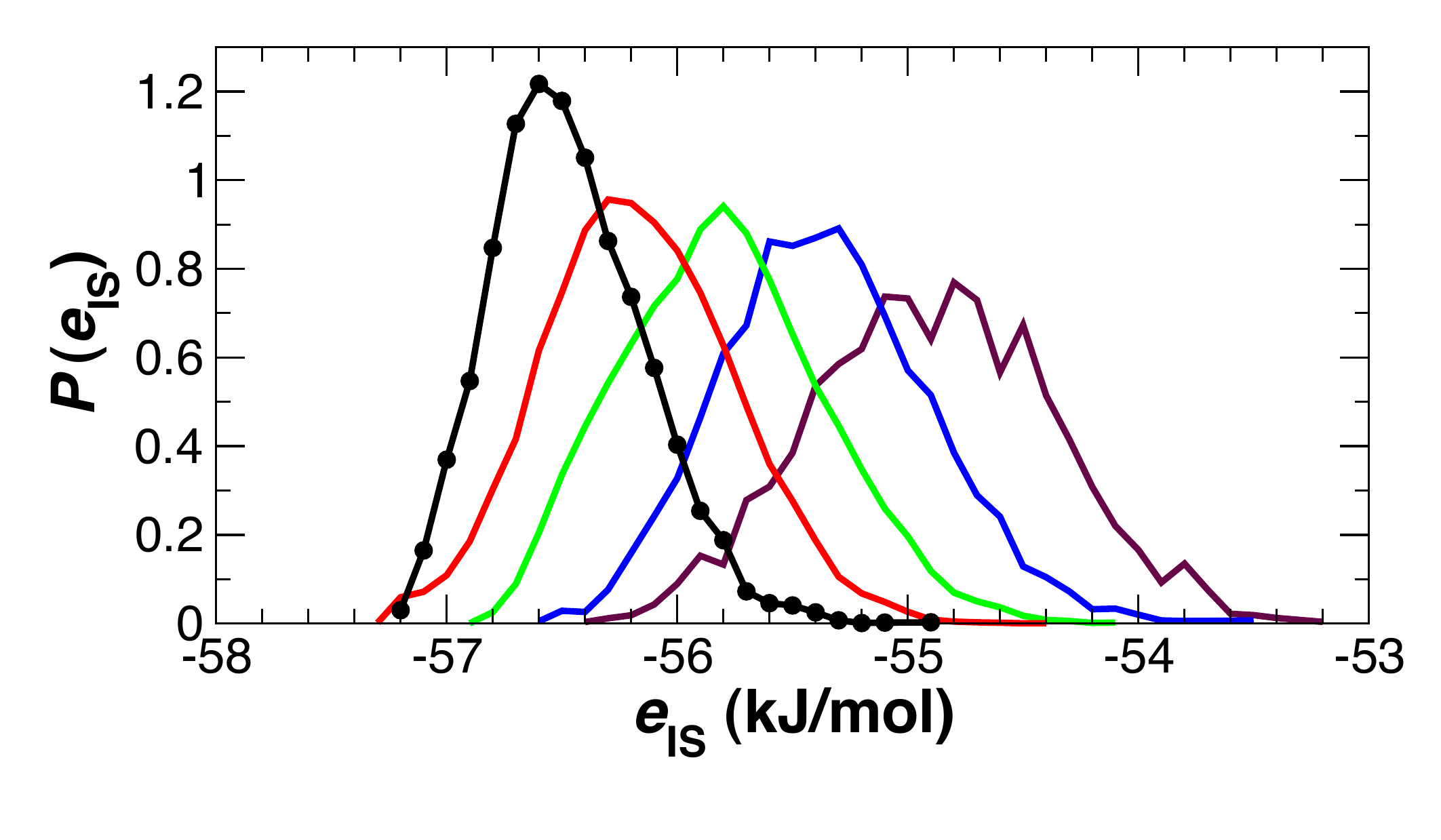}
\caption{Probability distributions of $e_{\rm IS}$ for the liquid from
  the $N=216$ simulations, along the $\rho=0.83$~g/cm$^3$ isochore.
  From left to right, $T=255$ to $275$~K, in 5~K steps.   }
\label{fig:Peis}
\end{figure}

The approach of the liquid to the bottom of the PEL is also confirmed in
Fig.~\ref{fig:Peis}, which plots the sampled probability distribution of the set of
$e_{\rm IS}$ values obtained at low $T$.  While at high $T$ the
distributions are approximately Gaussian, at the lowest $T=255$~K the
distribution has become distinctly skewed and narrower, reflecting the
approach to a finite lower bound for the possible values of $e_{\rm
  IS}$.  We assume that the approach of $e_{\rm IS}$ toward its minimum
value $e^0_{\rm IS}$ obeys  
\begin{equation}
e_{\rm IS}= e^0_{\rm IS} + B e^{-A_{ IS} /RT}
\end{equation}
and estimate $e^0_{\rm IS}$ by fitting $e_{\rm IS}$ to this form over
the range of $T$ in which $D$ appears to follow an Arrhenius behavior.
We approximate this range as the lowest four $T$ shown in Fig.~\ref{fig:Dfe},
and obtain $e^0_{\rm IS}=-57.6$~kJ/mol.  As expected for a liquid, this
value is above that of the value of $U$ for the crystal, ice~I$_h$.  The
limiting strong behavior will only be apparent when the system nears
$e^0_{\rm IS}$.  We notice that the value of $e^0_{\rm IS}$ is almost
exactly half of the value of the activation energy for diffusion,
supporting the view that $D$ is strongly related to the single particle
energy.  Since each bond connects two molecules, we can evaluate the
binding energy per bond $E_b \approx e^0_{\rm IS}/2 =-28.8$~kJ/mol,
which is larger than the bare HB energy $E_{\rm HB} \approx 20$~kJ/mol~\cite{sf89,sgs92}, since $E_b$ includes non-bonded
interactions.

We also note in Fig.~\ref{fig:huge}(b) that a maximum in the isochoric heat
capacity $C_V$ occurs in the vicinity of $T=285$~K, and as expected,
coincides with the inflections of $U$ and $e_{\rm IS}$ shown in
Fig.~\ref{fig:huge}(a).  This thermodynamic signature emphasizes that even
though all our data at $\rho=0.83$~g/cm$^3$ lie above $T_c$, for
$T<285$~K we have crossed into the regime in which the properties of the
liquid are increasingly dominated by the LDL phase, which is distinct
from the HDL phase for $T<T_c$.

\begin{figure}[!t]
\includegraphics[width=12cm]{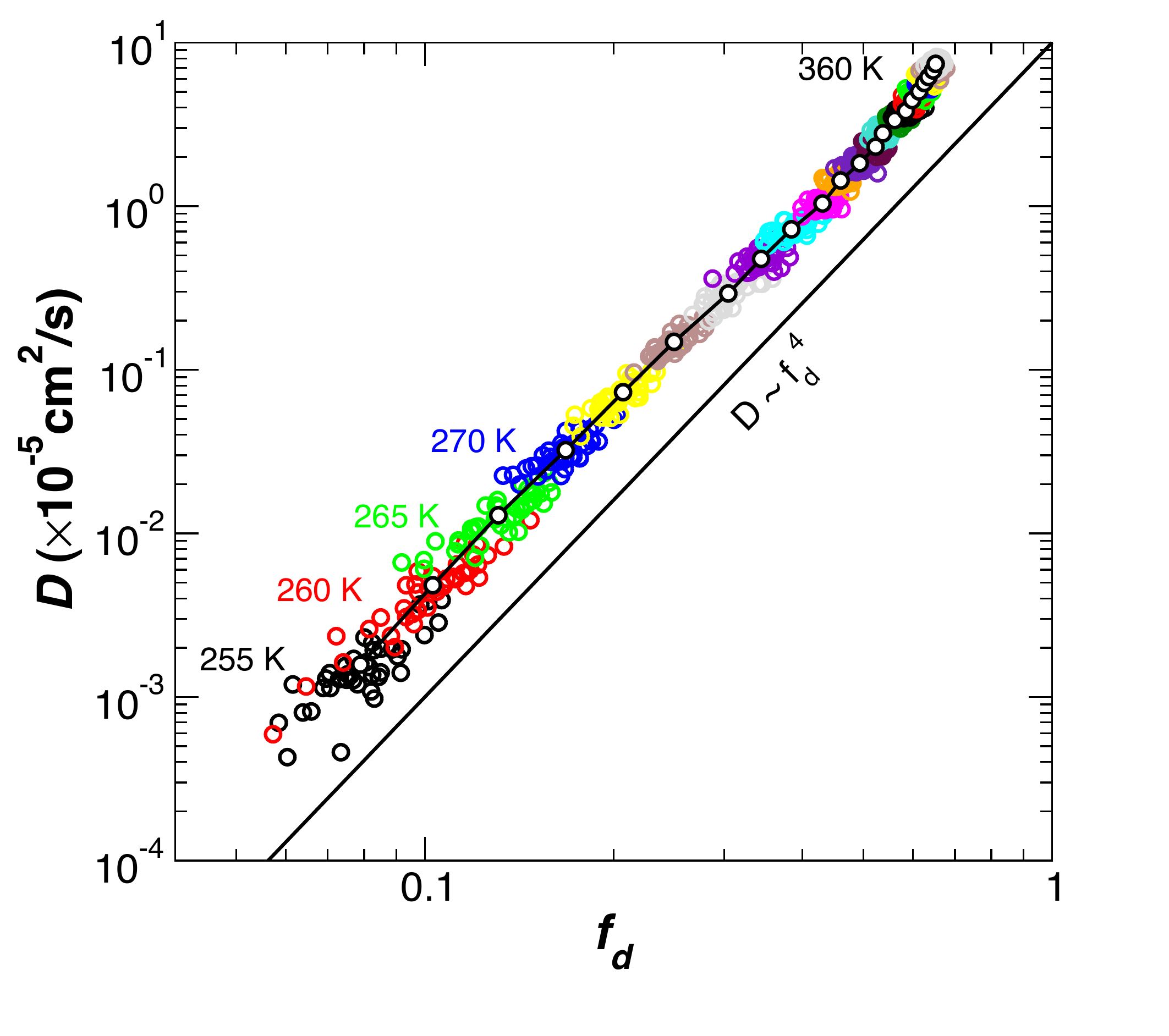}
\caption{Parametric plot of $D$ versus $f_d$ (squares) for the liquid
  from the $N=216$ simulations, along the $\rho=0.83$~g/cm$^3$ isochore,
  from $T=255$K (bottom left) to $360$~K (top right).  In this plot,
  $f_d$ is evaluated from the instantaneous configurations of the
  liquid.  For each $T$, we also show the ``cloud'' of points giving the
  values of $D$ and $f_d$ obtained from each of the 40 independent
  simulations conducted at each $T$.  The straight line has a slope of
  4.}
\label{fig:Df}
\end{figure}

In order to provide some structural insight into the dynamical behavior
along the $\rho=0.83$~g/cm$^3$ isochore, we examine the role of defects
in the RTN structure of the liquid.  In a perfect RTN, all molecules
would have exactly four nearest neighbors (nn's) within a distance given
by the first minimum of the oxygen-oxygen radial distribution function.
This distance is approximately 0.35~nm in ST2 water at
$\rho=0.83$~g/cm$^3$~\cite{sgs92}.  We thus define the fraction of RTN
defects $f_d$ as the average fraction of molecules in the system that do
{\it not} have four nn's within a distance of 0.35~nm; hence such defect
molecules will either have more than, or less than, four nn's.
$f_d$ is plotted as a function of $T$ in Fig.~\ref{fig:huge}(c), as obtained
from both instantaneous liquid configurations, as well as from inherent
structures.  In both cases we find that $f_d$ decreases as $T$
decreases, passes through an inflection near $T=285$~K, and approaches
zero at our lowest $T$. Indeed, as shown in Fig.~\ref{fig:Dfe}, we find that
the approach of $f_d$ to zero at low $T$ closely follows an exponential
decay, and in particular, that the defect activation energy $E_a$
obtained by plotting $f_d^4$ is approximately equal to the value of
$E_a$ obtained by fitting $D$ to an Arrhenius law at low $T$.  Since
$E_a\approx 4 E_b$, $f_d$ is approximately proportional to the exponential
of $E_b/RT$; hence, $f_d$ essentially measures the probability that a
single bond is broken.

To test if the relation between $D$ and $f_d$ holds at higher $T$ (where
the behavior is non-Arrhenius) we present in Fig.~\ref{fig:Df} a parametric
plot of $D$ versus $f_d$ over the entire $T$ range studied at
$\rho=0.83$~g/cm$^3$, from $T=360$~K to 255~K.  Except for some
deviations at the highest $T$, we find a remarkably consistent behavior
in which $D$ varies as a power law in $f_d$ over almost four decades,
and with an exponent very close to 4.  This result strongly suggests
that the liquid at this density can be understood as a disrupted RTN,
and that the localized excitations of this RTN ({\it i.e.}\ the defects)
control the transport properties of the liquid.

The relation between $D$ and $f_d$ can be anticipated by works on
colloids and nanoparticles with highly directional interactions.  In
particular, it was found that in systems of patchy colloids with four
sticky-spots in a tetrahedral geometry\cite{pwmnoi} and in nanoparticles
linked by DNA with tetrahedral orientation\cite{dnastarr,lss} $D$ is
given by the fourth power of the fraction of broken bonds, $D\sim f^4$.
One possible explanation for this behavior is that $D$ is controlled by
the diffusion process of the particles which have four broken bonds, and
which are thus free to wander around in the available empty space,
searching for the rare free dangling ends --- {\it i.e.}\ for the rare
sites of the network with incomplete bonding --- to stick.  Indeed, if
this is the case, $D$ is proportional in a first approximation to the
fraction of un-bonded particles, and this fraction scales with $f^4$.
It is important to remember that liquid water at this density
belongs\cite{statphys} to the category of so-called empty
liquids\cite{BianchiPRL}, {\it i.e.} liquids in which the fraction of
space occupied by a space-filling representation of the molecules is
significantly smaller than the close packing value. For the case of
liquid water, the corresponding effective packing fraction, when
molecules are considered as hard-spheres of diameter $0.28$~nm (the mean
hydrogen bond length), is $38 \%$.

The comparison between water and these tailored tetrahedral
systems~\cite{pwmnoi,dnastarr,lss} can also help clarify the connection
between the activation energy of the dynamics and the bonding
energy. Indeed, in these systems, it is clear that the activation energy
for diffusion is given by four times the bond energy, since the
interactions are short-ranged and there is no energy contribution from
non-bonded neighbors.  For ST2, each hydrogen bond has a strength of
$E_{\rm HB} \approx 20$~kJ/mol~\cite{sf89,sgs92}, significantly less (by
9~kJ/mol) than the overall binding energy per bond $E_b$ at the lowest
$T$ simulated.  If breaking of HBs are the limiting factor in diffusion,
we can anticipate that the asymptotic activation energy for diffusion
may decrease to a limiting value of $\approx 80$~kJ/mol, similar to the
low-$T$ value (74~kJ/mol) estimated for water~\cite{sas}.

\section{Summary}

In conclusion, we have provided a comprehensive survey of the diffusive
properties of the ST2 water model in the vicinity of the liquid-liquid
transition.  Our results demonstrate that the structural and dynamical
properties of the LDL phase that becomes a thermodynamically distinct
liquid phase for $T<T_c$ are already well established in the liquid at
low density for $T>T_c$.  The LDL phase is here revealed as a highly
structured liquid, whose properties are dominated by the progressive
emergence of a RTN structure as $T$ decreases.  In this sense, the
properties of the LDL phase are entirely consistent with those of low
density amorphous ice, as determined in experiments. Finally, we show
that the dynamics of the LDL appears to be fully controlled by the
presence of defects of the network, whose concentration is controlled by
the bond energy.

\section{Acknowledgments}

PHP thanks NSERC and the CRC Program for financial support, and ACEnet
for computational resources. FS acknowledges support from grant
ERC-226207-PATCHYCOLLOIDS.  FWS thanks the NSF for support from grant
number CNS-0959856.



\providecommand*\mcitethebibliography{\thebibliography}
\csname @ifundefined\endcsname{endmcitethebibliography}
  {\let\endmcitethebibliography\endthebibliography}{}

\end{document}